\documentclass[11pt]{article}
\usepackage[hyperref]{ccl2021-en}
\usepackage{times}
\usepackage{url}
\usepackage{latexsym}
\usepackage{fancyhdr}
\usepackage{graphicx}
\usepackage{pgfplots}
\usepackage{multirow}
\usepackage{amsmath}
\title{Using Query Expansion in Manifold Ranking for Query-Oriented Multi-Document Summarization}

\author{Quanye Jia \\
	Beihang University  \\
	{\tt jiaquanye@buaa.edu.cn} \\\And
	Rui Liu \\
	Beihang University \\
	{\tt lr@buaa.edu.cn} \\\And
	Jianying Lin \\
	Beihang University  \\
	{\tt jianying.lin@buaa.edu.cn} \\
}

\date{}

\begin{document}
\maketitle
\begin{abstract}
 Manifold ranking has been successfully applied in query-oriented multi-document summarization. It not only makes use of the relationships among the sentences, but also the relationships between the given query and the sentences. However, the information of original query is often insufficient. So we present a query expansion method, which is combined in the manifold ranking to resolve this problem. Our method not only utilizes the information of the query term itself and the knowledge base WordNet to expand it by synonyms, but also uses the information of the document set itself to expand the query in various ways (mean expansion, variance expansion and TextRank expansion). Compared with the previous query expansion methods, our method combines multiple query expansion methods to better represent query information, and at the same time, it makes a useful attempt on manifold ranking. In addition, we use the degree of word overlap and the proximity between words to calculate the similarity between sentences. We performed experiments on the datasets of DUC 2006 and DUC2007, and the evaluation results show that the proposed query expansion method can significantly improve the system performance and make our system comparable to the state-of-the-art systems.
\end{abstract}

\section{Introduction}
\label{intro}

%
%
\cclfootnote{
    %
    %
    \hspace{-0.65cm}  
    \textcopyright 2021 China National Conference on Computational Linguistics

    \noindent Published under Creative Commons Attribution 4.0 International License
}

Query-focused multi-document summarization is to create a summary from a set of documents that answers the information requirements expressed in the query. Compared to generic summarization, query-focused summarization requires the summary biased to a specific query besides the general requirement for a summary. In contrast to the task of question answering (QA) that mainly focuses on simple factoid questions and results in precise answers such as person, location or date, etc., in the case of query-focused summarization, the queries are mostly real-world complex questions and the information provided by the queries is insufficient. So how to understand the query and expand the query has a certain impact on the quality of the query-oriented summary.

Currently most query oriented abstract systems are based on general summarization systems. By incorporating features that are related to the given query (e.g., the relevance of a sentence to the query), a generic summarization system can be adapted to a query-focused one. Due to the limit of information that a query can express, some systems also expand the query using some external resources such as WordNet, by which the synonyms of the query words can be obtained as expansion words \cite{zhou2005based}. However, such approaches to query expansion are restricted in query itself and external resources, because they cannot be applied to words not in WordNet such as named entities which frequently occur in queries, and much context related information cannot be captured by only synonyms. Lin\cite{Lin2009Using} uses graph-based sentence ranking and sentence-to-word relationships to implement query expansion, which can improve the effect of graph ranking. However, the paper uses a single query expansion method and only adopts TextRank instead of manifold ranking as graph ranking algorithm. So we improved the query expansion method based on the above problems.

The main contributions of this work are:

\begin{itemize}
	\item We used multiple query expansion methods to expand the query, including the query itself (using Wordnet synonyms) and the expanded query based on the document set (mean expansion, variance expansion and TextRank expansion).
	\item We improved the calculation of sentence similarity matrix in the manifold ranking process, using sentence TF-ISF similarity, word overlap similarity, and sentence proximity information.
\end{itemize}

Our experiments on the dataset DUC 2006 and DUC 2007 show that the summarization results with query expansion are much better than that without query expansion, achieving the state-of-the-art performance.

\section{Related Work}
Graph-based ranking algorithms have been successfully used in text summarization. \cite{ErkanLexRank2004} proposed LexRank for generic text summarization. They construct a connected similarity graph where nodes represent sentences and edges represent cosine similarities between sentences. A random walk is applied on the graph until converging to a stationary distribution, by which the sentences can be ranked. Topic-sensitive LexRank\cite{Otterbacher2005using} has been applied to the task of query-focused summarization where the relevance of a sentence to the query is taken into account when performing random walk. \cite{Xiong2016Query} proposed a hypergraph based vertex-reinforced random walk framework for multi-document summarization. \cite{Wan2007Manifold} applied a manifold ranking algorithm to query-focused summarization which can simultaneously make full use of both the relationships among all the sentences in the documents and the relationships between the given query and the sentences.

Regarding manifold ranking algorithms, subsequent researchers have made a lot of improvements. \cite{Tan2015Joint} introduced the matrix factorization model on the basis of Xiaojun Wan's model, and constructed the sentence relational matrix by using the cosine value of the decomposed matrix. \cite{Chali2011Using} used the syntactic and shallow semantic kernels to calculate the correlation between sentences based on the model of Xiaojun Wan.  On the basis of Xiaojun Wan's model, \cite{Lin2019Sum} improves the relationship matrix by mixing the cosine similarity of decomposed matrix and TF-ISF cosine similarity in a certain proportion. At the same time, this paper uses lifelong learning topic model to enhance the co-occurrence effect of matrix factorization words. Finally, this paper adds statistical features, mix various scoring methods together, and then extract summaries with the strategy of de-redundancy. \cite{2012Mani} not only considers the internal relevance propagation within the sentence set (or within the theme cluster set), but also considers the mutual reinforcement between the ranking of the sentence set and the ranking of the cluster set. These models mainly focus on the calculation of sentence similarity and sentence ranking, but pay little attention to queries.

Query understanding is a research hotspot in the field of search engines. \cite{Abdelali2007Improving} proposed the use of a corpus to expand the original query. \cite{Jian2009Understanding} proposed a Wikipedia-based user query intent understanding. \cite{Pal2014Improving} considered the distribution of expansions in documents, statistical associations with query terms, and semantic associations.\cite{Wang2011Semantic} and \cite{Pinto2009Joining} combined query expansion with word meaning elimination to enhance the validity of the query. To the best of our knowledge, few people have done the research on the effect of query understanding and query expansion on manifold ranking. Therefore, based on the above research, we use multiple query expansion methods to improve the effect of model based on manifold ranking.

\section{Method}
\begin{figure*}
	\includegraphics[width=\textwidth]{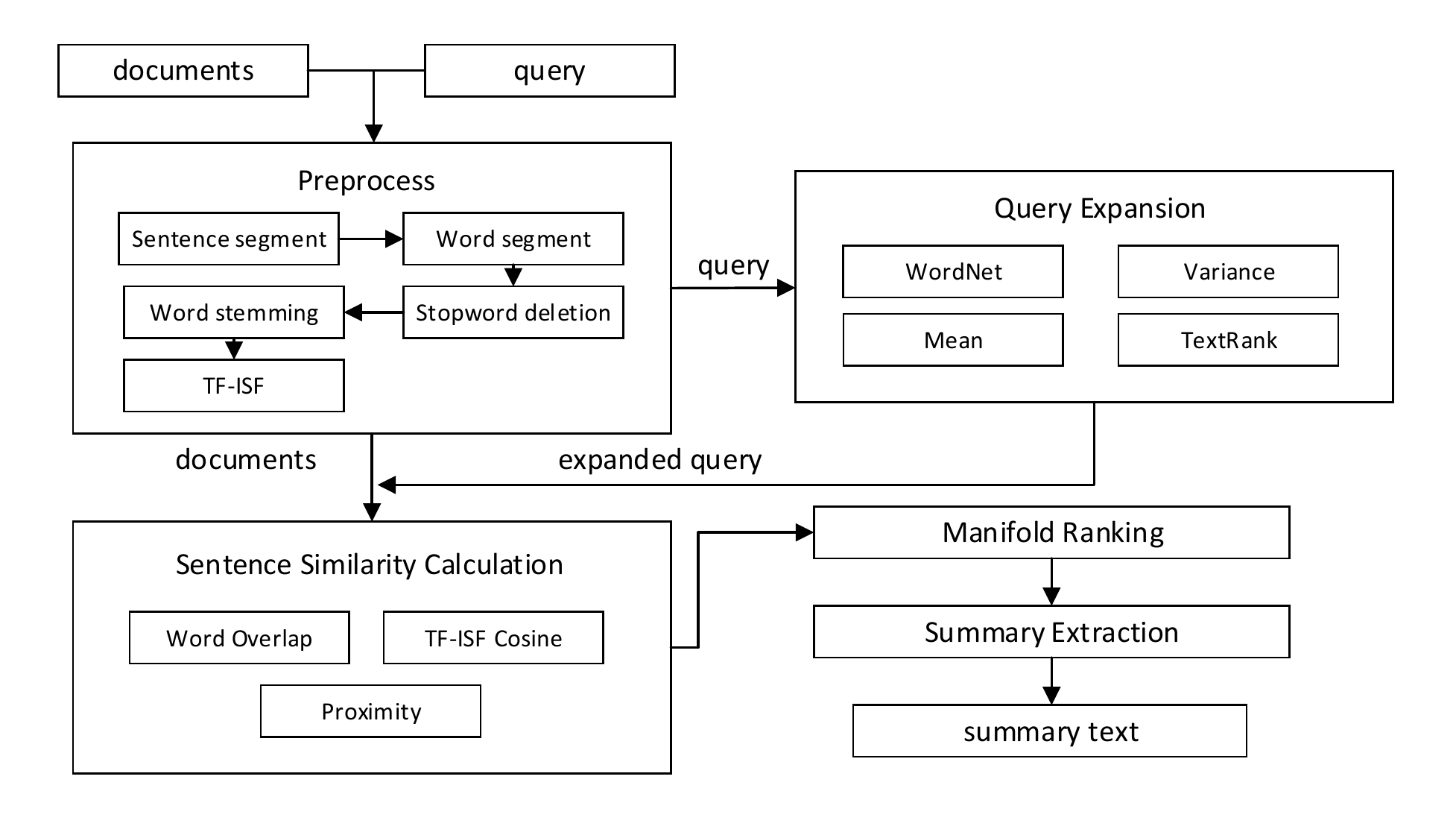}
	\caption{Model architecture.} \label{fig:architecture}
\end{figure*}

In this section, this paper will introduce the model architecture, as Fig.\ref{fig:architecture} shows.

For the preprocessing step, this paper performed sentence segmentation, word segmentation, word stemming and TF-ISF calculation for each sentence in the documents and the query. This paper utilized titles and narratives as queries. When computing the sentence similarities, stop words, emails, digital characters, punctuation marks and single characters were removed, and only the content words (i.e., noun, verb, adjective and adverb) were used after the morphological analysis based on treetagger.

For the extraction process of automatic text summarization, this paper selects the method used by \cite{Wan2007Manifold}.

Manifold Ranking, query expansion and sentence similarity calculation and will be introduced in detail in following sections.  

\subsection{Manifold Ranking}
Manifold ranking \cite{Zhou2004Learning} \cite{zhou2004ranking} is a universal graph ranking algorithm and it is initially used to rank data points along their underlying manifold structure. The prior assumption of manifold-ranking is: (1) sentences on the same structure are likely to have the same ranking scores; (2) sentences similar to the query sentence have higher ranking scores. The process of the manifold ranking algorithm can be understood as follows: construct a similarity matrix W between all sentences, initialize a vector $y=[y_{0},y_{1},\cdots,y_{n}]^T$,  in which $y_{0}=1$ because $x_{0}$ is the query sentence and $y_{i}=0 (1 \leq i \leq n)$ for all the sentences in the documents. Then the ranking algorithm transmits the scores of query sentences to the scores of sentences similar to query points through the similarity between sentences, and then the scores of each sentence are propagated to the scores of other similar sentences, iterating many times until the scores of each sentence remain unchanged. The algorithm for manifold ranking is as Formula \ref{equ:mani} shows:

\begin{equation*}
\begin{split}
{\rm{L(}}f{\rm{)}} = {\alpha _{\rm{mr}}}*{\sum\limits_{i,j = 1}^N {W_{ij}\left| {\frac{1}{{\sqrt {D_{ii}} }}{f_i} - \frac{1}{{\sqrt {{\rm{D}_{\rm{jj}}}} }}{f_j}} \right|} ^2} + (1 - {\alpha _{mr}})*{\sum\limits_{i = 1}^N {\left| {{f_i} - {y_i}} \right|} ^2} 
\end{split}
\end{equation*}

\begin{equation}
\label{equ:mani}
\begin{split}
&{\rm{s}}{\rm{.t}}{\rm{.}}{D_{ii}}{\rm{ = sum(}}{{\rm{W}}_{i*}}{\rm{),}}Y =  \left[\, 
{y_1} \, {y_2} \, \cdots \, {y_N}
\right]^{\rm{T}} = {\left[\, 
	1 \, 0 \, \cdots \, 0 \, \right]^{\rm{T}}}
\end{split}
\end{equation}

Where $ F = {\left[\,{f_1} \quad {f_2} \quad \cdots \quad {f_n} \, \right]^T} $ is the sentence score vector for the solution, and $Y$ is the vector indicating the query sentence. This paper assumes that when ${A_i}$ is the query sentence or the query sentence ${y_i} = 1$, and otherwise is 0, ${W_{ij}}$ indicates the relationship between the i-th sentence and the j-th sentence. There are many ways to calculate the relationship between sentences, such as Manhattan distance, Euler distance, inner product calculation method, cosine similarity, etc., or a combination of various methods. ${D_{ij}}$ represents the sum of the elements of the i-th row in the W matrix, and ${\alpha _{{\rm{mr}}}}$ represents the proportion of the similarity between the sentences in the manifold ranking. 

\subsection{Query Expansion}
In the query-oriented multi-document summarization, each topic has one or two sentences as the query, but the information of these sentences is often insufficient. Therefore, the query sentence should be expanded. There are two general methods: one is to expand according to the content of documents. The other is to expand based on the query itself, which usually depends on the understanding of the query. For each topic, we extract query sentences and sentences of multiple documents to form a sentence-word matrix ${A^{(0)}} = {\left[ \; {T^{(0)}} \quad X \; \right]^{\rm{T}}}$ of N*M, and then use the above two methods to expand the query sentence, the formula for expanding query sentences is as Formula \ref{equ:ext} shows:

\begin{equation}
\label{equ:ext}
\begin{split}
{T^{(1)}}={\rm{sim}}\_{\rm{word}}({T^{(0)}}) + {\theta _m} {\rm{mean}}(X) + {\theta _v} {\rm{variance}}(X) + {\theta _r} {\rm{TextRank}}(A^{(0)})
\end{split}
\end{equation}

Where $A$ is a sentence-word matrix, the word weight is calculated using the TF-ISF(as \cite{Lin2009Using} use) formula, $T$ is query sentences, $X$ is all document sentences, N is the number of the sentences, M is the number of all various content words in the documents, ${\rm{sim}}\_{\rm{word}}({T^{(0)}})$ is the expansion of the word meaning of the query sentence, ${\rm{mean}}(X)$ is the mean expansion of the query sentence, ${\rm{variance}}(X)$ is the variance expansion of the query sentence, ${\rm{TextRank}}(A^{(0)})$ is the TextRank expansion of the query sentence.

\subsubsection{Query Expansion By WordNet}
WordNet is a large English word sense database\cite{WordNet}. It not only explains the meaning of words, but also links the meaning of words together. In WordNet, each of synset(semantics) represents a basic lexical concept. A word can have more than one synonym, and a synonym can correspond to more than one word. Synonymous phrases are connected by semantic relations such as hypernymy-hyponymy, synonym-antonym, meronymy-holonyms and implication, forming a network structure with synonymous phrases as nodes and semantic relations as edges. Therefore, semantic similarity can be calculated by the length of the edge between two semantic nodes. 

Semantic similarity calculation is based on \cite{Rada1989Rada}, which mainly uses the path length between concepts. The shorter the path between two nodes is, the higher the semantic similarity between the two concepts is. In this paper, the similarity is calculated using the following formula \ref{equ:sim}:

\begin{equation}
\label{equ:sim}
{\rm{sim}}(syn_1,syn_2) = \frac{a}{{a + d}}
\end{equation}

Where $syn1$, $syn2$ refers to the semantic node of a word, d represents the number of upper edges of the shortest path between the semantic nodes $syn1$ and $syn2$ in the semantic graph, and a represents an adjustable parameter.

For most applications, there is no data with word meaning labels, so algorithms are needed to provide us with similarities between words rather than between meanings or concepts. For any algorithm based on a semantic dictionary, according to Resnik \cite{Resnik1995Using}, we can approximate the similarity by using the maximum similarity between two word semantic items. Therefore, based on word meaning similarity, we can define word similarity as following formula \ref{equ:max_sim}:

\begin{equation}
\label{equ:max_sim}
sim({w_1},{w_2}) = \mathop {\max }\limits_{\scriptstyle{syn_1} \in senses({w_1})\hfill\atop
	\scriptstyle{syn_2} \in senses({w_2})\hfill} {\rm{sim}}({syn_1},{syn_2})
\end{equation}

Where $w1$, $w2$ represent the target word, $syn1$, $syn2$ refer to the sense of the word. For a sentence-word matrix ${A^{(0)}}$ of N*M, the similarity of all words in the document can be obtained first, and then the word similarity matrix of M*M can be obtained. According to the matrix, each word of the query sentence can be expanded. The expanded formula is as Formula \ref{equ:ext_sim_word} shows.

\begin{equation}
\label{equ:ext_sim_word}
t_i^{(1)} = \mathop {\max }\limits_j (t_j^{(0)} \cdot {\rm{sim}}({w_j},{w_i}))
\end{equation}

$t_i^{(0)}$ represents the i-th word feature before expansion, $t_i^{(1)}$ represents the expanded i-th word feature, and ${w_i}$ represents the i-th word item, then ${{\rm{T}}^{(1)}}={\rm{sim}}\_{\rm{word}}({T^{(0)}}) = \left[ \; {t_1^{(1)}} \quad {t_2^{(1)}} \quad \cdots \quad {t_M^{(1)}} \; \right]$ is the query sentence after the word meaning expansion. By expanding the meaning of the query sentence, it can make the query sentence reflect the features of the similar words in the original sentence, and solve the problem that the feature of the query sentence is not salient because the original query sentence has too few words.

\subsubsection{Word Similarity Filtering}
According to the above word similarity calculation method, we can extract all words from the dataset to calculate the word similarity matrix. However, most of the similarities between words have no practical significance, so they need to be filtered. There are two kinds of filtering methods in this paper: vertical filtering and horizontal filtering.

Vertical filtering is based on the path length between words. For example, computer and football are obviously not similar, but because they can be connected by superordinate entities, even if the shortest path between them is very long, they will still get smaller values after similarity calculation. This degree of similarity has little significance for feature calculation, so smaller values can be discarded. In this way, not only computational meaningful features can be expanded, but also sparse matrix storage can be used to reduce the waste of storage space. Therefore, this paper will use vertical filtering parameter as L to filter the similarities which the shortest path is greater than L.  

Horizontal filtering is based on the number of similarities between a word and other words. When a word is similar to most words, it shows that the feature of the word itself is not salient, so it needs to be filtered. For example, the verbs "use" and "do" are similar to almost all verbs. If these verbs are expanded, noise will be introduced into the textual features of the query sentences, which will not highlight the topic feature of the query sentences.  Therefore, this paper will use horizontal filtering parameter as C to retain the words which are similar to no more than C other words.

\subsubsection{Query expansion By Mean and Variance}
Due to the large amount of information contained in the text, the average TF-ISF value and the variance TF-ISF value of each word in the document can be obtained to expand the query sentence, as shown in the following Formula \ref{equ:ext_mean} and \ref{equ:ext_var} show.

\begin{equation}
\label{equ:ext_mean}
\begin{split}
{\rm{mean}}(X) = \left[ {\bar X}_{*1} \quad {...} \quad {\bar X}_{*M}\right] = \left[\
{\frac{1}{{N - 1}}\sum\limits_{i = 1}^{N - 1} {X_{i1}} } \quad {...} \quad {\frac{1}{{N - 1}}\sum\limits_{i = 1}^{N - 1} {X_{iM}} }\right]
\end{split}
\end{equation}

\begin{equation}
\label{equ:ext_var}
\begin{split}
{\rm{variance(}}X{\rm{) = }}\left[ 
{\frac{\rm{1}}{{\rm{N-2}}}\sum\limits_{\rm{i = 1}}^{\rm{N - 1}} {{{\rm{(}}{{\rm{X}}_{{\rm{i1}}}}{\rm{ - }}{{{\rm{\bar X}}}_{{\rm{*1}}}}{\rm{)}}}^{\rm{2}}} } \right. 
\left.{\frac{\rm{1}}{{\rm{N-2}}}\sum\limits_{\rm{i = 1}}^{\rm{N - 1}} {{{\rm{(}}{{\rm{X}}_{{\rm{i2}}}}{\rm{ - }}{{{\rm{\bar X}}}_{{\rm{*2}}}}{\rm{)}}}^{\rm{2}}} } \, {...}  \,  {\frac{{\rm{1}}}{{{\rm{N-2}}}}\sum\limits_{{\rm{i = 1}}}^{{\rm{N-1}}} {{{{\rm{(}}{{\rm{X}}_{{\rm{iM}}}}{\rm{ - }}{{{\rm{\bar X}}}_{{\rm{*M}}}}{\rm{)}}}^{\rm{2}}}} } \right]
\end{split}
\end{equation}

$N$ represents the number of all sentences, ${X_{ij}}$ represents the i-th document sentence and the j-th word item feature, ${\bar X_{*j}}$ represents the average of the j-th word item feature in the documents.

\subsubsection{Query expansion By TextRank}
In this section, we select from the document set both informative and query relevant words based on TextRank (section 3.2 in \cite{Lin2009Using}) results, add them into the original query and use the updated query to perform manifold ranking again. Our query expansion approach goes as follows:

1. Normalize $X$ by $S = D^{-1}X$ to make the sum of each row equal to 1, where $D$ is the diagonal matrix with (i, i)-element equal to the sum of the ith row of $X$.

2. Calculate vector $y$ by $y = S^{T}p^{*}$, where $p^{*}$ is the vector of sentence ranking scores derived in the last step in the TextRank algorithm described in Lin et al.\cite{Lin2009Using}, and y represents the word scores.

3. Rank all the words based on their scores in $y$ and select the top c words as query expansions. c is a parameter representing the number of expansion words, which is set in the experiments.

4. Add the top c words into the query sentence $T$, the formula is as follows:
\begin{equation}
TextRank(A^{(0)})=\left[t_{1}^{'},\quad t_{2}^{'},\quad {...},\quad t_{n}^{'} \right]
\end{equation}
\begin{equation}
t_{i}^{'}=\left\{ \begin{matrix}
\begin{matrix}
1 & \begin{matrix}
\text{when the words corresponding to t are in the top c words}  \\
\end{matrix}\text{ }  \\
\end{matrix}  \\
\begin{matrix}
0 & \begin{matrix}
\text{when the words corresponding to t are not in the top c words}  \\
\end{matrix}  \\
\end{matrix}  \\
\end{matrix} \right.
\end{equation}

Where ${t}_{i}^{'}$ represents the i-th word feature after expansion. This algorithm is to make use of both the sentence importance and sentence-to-word relationships to select the expansion words. By this step, salient words occurring in the important sentences are more likely to be selected, and because the higher ranked sentences are biased towards the query, the words selected in this way are also biased towards the query. 

\subsection{Sentence Similarity Calculation}
The W matrix is used to measure the similarity between two sentences, which is critical in manifold ranking. If the calculation method of W is closer to the similarity between sentences, the quality of the summary is higher. This paper mainly uses TF-ISF cosine similarity, word overlap and proximity similarity to calculate. The TF-ISF cosine similarity and word overlap use the word frequency characteristics of the sentence itself. Word overlap only considers the number of words co-occurring in two sentences, regardless of the occurrence of words in other sentences, which is beneficial to increase the number of different words in the summary. The adjacency similarity mainly considers the relationship between the relative positions of two sentences, so that the meaning of the sentence in the context is reflected. The three are organically combined, as shown in the following Formula \ref{equ:w}:

\begin{equation}
\label{equ:w}
{W_{ij}} = {\alpha _A} \cdot \cos ({A_{i{\rm{*}}}^{(1)}},{A_{j{\rm{*}}}^{(1)}}) + {\alpha _{overlap}} \cdot S{S_{ij}}{\rm{ + }}{\alpha _{p{\rm{eer}}}} \cdot {P_{ij}}
\end{equation}

Where $\cos ({A_{i{\rm{*}}}^{(1)}},{A_{j{\rm{*}}}^{(1)}})$ represents the cosine similarity of the TF-ISF of the sentence ${A_{i*}^{(1)}}$ and ${A_{j*}^{(1)}}$. $SS_{ij}$ indicates the degree of word overlap of the sentence ${A_{i*}^{(0)}}$ and ${A_{j*}^{(0)}}$, that is, the ratio of the number of words appearing together in two sentences to the minimum word length of two sentences.

For the position information of the sentence, this paper mainly considers the context of the sentence. $P_{ij}$ is used to represent the adjacency matrix of the sentence. One topic has multiple documents, and one document has multiple sentences. These sentences constitute the context and the sentences are in a specific document. There is a specific context, each sentence must have a certain relationship with the adjacent sentence, so the adjacency matrix can be used to quantify the relationship between the sentences. The closer the sentences are, the higher the similarity score is. The representation is as shown in the following Formula \ref{equ:P}:

\begin{equation}
\label{equ:P}
{{\rm{P}}_{ij}}{\rm{ = }}\left\{ {\begin{array}{*{20}{c}}
	{{{0.1}^{\left| {{\rm{ i  -  j }}} \right|}}, {\rm{i\;and\;j\;are\;in\;the\;same\;document}}}\\
	{0,\; {\rm{i\;and\;j\;are\;not\;in\;the\;same\;document}}}
	\end{array}} \right.
\end{equation}

In addition, when solving the cosine similarity matrix of TF-ISF between sentences, this paper uses the expanded query sentence as the query sentence. When calculating the sentence word overlap matrix, the original query sentence is used as the query sentence.

\section{Experimental Setup}
\subsection{Datasets and Evaluation}
The summary data sets used in this experiment are DUC2006\cite{DUC2006} and DUC2007\cite{DUC2007}. DUC 2006 and DUC 2007 are query-oriented multi-document summary datasets, each data set contains multiple topics, each topic consists of multiple related documents, and each topic provides a title and a narrative as a query. For each topic, we take the first 250 words as the model summary from the results and compare it with the expert summary by using ROUGE\cite{Lin2003ROUGE} toolkit. We report F1 scores of ROUGE-1, ROUGE-2, ROUGE-W and ROUGE-SU4 metrics.

\subsection{Parameter settings}
The parameters of the proposed methods are determined according to the overall effect of the model on DUC data set. The parameter a is 1, L is 4, C is 5000, the mean parameter ${\theta _m}$ is 1, the variance parameter ${\theta _v}$ is 1, ${r_t}$ is 0.4, the ${\alpha _{mr}}$ is 1, the cosine similarity parameter ${\alpha _A}$ is 0.9, the word coverage parameter ${\alpha _{overlap}}$ is 0.1, and the adjacency similarity parameter ${\alpha _{peer}}$ is 0.4. $\omega $ as a de-redundancy parameter, takes 8 (refer to \cite{Wan2007Manifold}). In the TextRank expansion, the c is 100 and d is 0.6 (refer to \cite{Lin2009Using}).

\section{Results}
\subsection{Comparison with query expansion methods}
In this experiment, our aim is to examine the efficiency of the combination of these query expansion methods. The results are reported in table \ref{tab:ext}.

\begin{table*}[t]
	\centering  
	\begin{tabular}{|l|l|l|l|l|l|}
		\hline
		{Dataset}&{Extexsion Method} &{Rouge-1}&{Rouge-2}&{Rouge-W} &{Rouge-SU4} \\
		\hline		
		\multirow{11}{*}{DUC2006} & ori                 & 0.40331 & 0.08809 & 0.13960  & 0.14637 \\
		& ori+MEAN            & 0.40695 & 0.09057 & 0.14099  & 0.14827 \\
		& ori+VAR             & 0.40604 & 0.09013 & 0.14060  & 0.14835 \\
		& ori+SIM\_WORD       & 0.41019 & 0.08785 & 0.14084  & 0.14691 \\
		& ori+TextRank        & 0.41658 & 0.09174 & 0.14274  & \textbf{0.15102} \\
		& ori+SIM\_WORD+MEAN  & 0.41129 & 0.08895 & 0.14078  & 0.14762 \\
		& ori+SIM\_WORD+VAR   & 0.41112 & 0.08879 & 0.14113  & 0.14762 \\
		& ori+MEAN+VAR & & & &\\
		& +SIM\_WORD          & 0.41060 & 0.08934 & 0.14090  & 0.14723 \\
		& \emph{\textbf{ori+MEAN+VAR}} & & & &\\
		& \emph{\textbf{+SIM\_WORD+TextRank}} & \textbf{0.41674} & \textbf{0.09202} & \textbf{0.14279} & \textbf{0.15071} \\  
		\hline
		\multirow{11}{*}{DUC2007} & ori                    & 0.42010 & 0.10372 & 0.14559  & 0.16051  \\
		& ori+MEAN               & 0.42612 & 0.10638 & 0.14773  & 0.16329  \\
		& ori+VAR                & 0.42149 & 0.10444 & 0.14616  & 0.16102  \\
		& ori+SIM\_WORD          & 0.42565 & 0.10195 & 0.14564  & 0.15991  \\
		& ori+TextRank           & 0.43557 & 0.11174 & 0.15088  & 0.16668  \\
		& ori+SIM\_WORD+MEAN     & 0.42907 & 0.10447 & 0.14706  & 0.16163  \\
		& ori+SIM\_WORD+VAR      & 0.42699 & 0.10184 & 0.14604  & 0.16016  \\
		& ori+MEAN+VAR & & & &\\
		& +SIM\_WORD            & 0.43042 & 0.10574 & 0.14768  & 0.16295  \\
		& \emph{\textbf{ori+MEAN+VAR}} & & & &\\
		&\emph{\textbf{+SIM\_WORD+TextRank}} & \textbf{0.43982} & \textbf{0.11185} & \textbf{0.15159} & \textbf{0.16870}\\   
		\hline
	\end{tabular}
	\caption{ROUGE Evaluation Results of Expanded Query Sentences and Original Query Sentences in DUC2006 and DUC2007.\newline}
	\label{tab:ext} 
\end{table*}

From the report, we can figure out that the performance of the method using expansion of query is better than that using the  original query. In other words, the results of ori+MEAN+VAR+TextRank are better than one that combines a subset of query expansion methods. 

\begin{table*}[th]
	\centering  
	\begin{tabular}{|l|l|l|l|l|l|}
		\hline
		{Dataset}& {System} & {Rouge-1} & {Rouge-2} & {Rouge-W} & {Rouge-SU4} \\
		\hline		
		\multirow{16}{*} {DUC2006} & {Random} & {0.35352} & {0.05381} & {0.12135} & {0.111} \\
		&{Lead} & {0.3396} & {0.05412} & {0.11748} & {0.10961} \\
		&{MV-CNN} & {0.3865} & {0.0791} & {-} & {0.1409} \\
		&{QODE} & {0.4015} & {0.0928} & {-} & {0.1479} \\
		&{AttSum} & {0.409} & \textbf{0.094} & {-} & {-} \\
		&{VAEs-A} & {0.396} & {0.089} & {-} & {0.143} \\
		&{C-Attention} & {0.393} & {0.087} & {-} & {0.141} \\
		&{MultiMR} & {0.40306} & {0.08508} & {0.13997} & {-} \\
		&{SingleMR} & {0.39934} & {0.07502} & {0.13445} & {0.13404} \\
		&{JMFMR} & {0.41244} & {0.0887} & {-} & {0.14585} \\
		&{JTMMR} & {0.40014} & {0.08160} & {0.13652} & {0.14062} \\
		&{JLTMMR} & {0.4043} & {0.08126} & {0.13774} & {0.14032} \\
		&{JLTMMR+SF} & {0.41045} & {0.08926} & {0.14214} & {0.14798} \\
		&{RDRP\_AP} & {0.39615} & {0.08975} & {-} & {0.13905} \\
		&{TextRank} & {0.39056} & {0.0871} & {0.13536} & {0.14299} \\
		&{QE-T+TextRank} & {0.40115} & {0.08985} & {0.13804} & {0.14619} \\
		&\emph{\textbf{QE-WMVT+Mani}} & \textbf{0.41674} & {0.09202} & \textbf{0.14279} & \textbf{0.15071}\\
		\hline
		\multirow{16}{*}{DUC2007} & {Random} & {0.36896} & {0.06654} & {0.125} & {0.12312} \\
		&{Lead} & {0.35461} & {0.06639} & {0.1219} & {0.11926} \\
		&{Kmeans+NMF} & {0.37076} & {0.07194} & {0.12786} & {0.12814} \\
		&{MV-CNN} & {0.4092} & {0.0911} & {-} & {0.1534} \\
		&{QODE} & {0.4295} & \textbf{0.1163} & {-} & {0.1685} \\
		&{AttSum} & {0.4392} & {0.1155} & {-} & {-} \\
		&{VAEs-A} & {0.421} & {0.11} & {-} & {0.164} \\
		&{C-Attention} & {0.423} & {0.107} & {-} & {0.161} \\
		&{MultiMR} & {0.42041} & {0.10302} & {0.14595} & {-} \\
		&{SingleMR} & {0.41422} & {0.09052} & {0.13984} & {0.14589} \\
		&{JTMMR} & {0.42717} & {0.10181} & {0.1458} & {0.15761} \\
		&{JLTMMR} & {0.43349} & {0.10375} & {0.14786} & {0.16002} \\
		&{JLTMMR+SF} & {0.43734} & {0.10439} & {0.1496} & {0.1625} \\
		&{RDRP\_AP} & {0.43775} & {0.11563} & {-} & \textbf{0.16904} \\
		&{HERF} & {0.42444} & {0.11211} & {-} & {0.16307} \\
		&{TextRank} & {0.4033} & {0.10196} & {0.14063} & {0.15463} \\
		&{QE-T+TextRank} & {0.41554} & {0.10597} & {0.14379} & {0.16074} \\
		&\textbf{\textbf{QE-WMVT+Mani}} & \textbf{0.43982} & {0.11185} & \textbf{0.15159} & {0.16870}\\
		\hline			
	\end{tabular}
	\caption{Rouge results of multi-document summarization on DUC2006 and DUC2007.}
	\label{ROUGE Result 2006 and 2007} 
\end{table*}

\subsection{Comparison with related methods}
In this experiment, we will compare the model performance with the other existing well-known methods. We compare QE-WMVT-Mani(Manifold ranking with word similarity, mean, variance and TextRank query expansion) with other methods: 1)Random(For each topic, randomly extracts a certain number of sentences as a summary); 2)Lead(Under each topic, select a certain number of sentences in the document of the most recent time as a summary);  3)MV-CNN\cite{Zhang2016CNN}; 4)AttSum\cite{Cao2016AttSum}; 5)VAEs-A\cite{Li2017VAEs-A}; 6)QODE\cite{Liu2015QODE}; 7)C-Attention\cite{Li2017C-Attention}; 8)MultiMR\cite{Wan2009MultiMR}; 9)JMFMR\cite{DBLP:conf/sigir/TanWX15}; 10)JTMMR\cite{Lin2019Sum}; 11)JLTMMR\cite{Lin2019Sum}; 12)JLTMMR + SF\cite{Lin2019Sum};
13)RDRP\_AP\cite{2012Mani}; 14)HERF\cite{Xiong2016Query};  15)TextRank\cite{2004TextRank}; 16)QE-T+TextRank\cite{Lin2009Using}.

For MV-CNN, QODE, AttSum, VAEs-A, C-Attention, MultiMR, JMFMR, JTMMR, JLTMMR, JLTMMR+SF, RDRP\_AP and HERF models, this paper chooses results from their corresponding papers. For the QE-T+TextRank method, because the original paper does not give the F value, this paper re-experiments it, and the preprocessing process is the same as the original paper. ${r_t}$ is set to 0.4, and other parameters are the same as the original paper.

Table \ref{ROUGE Result 2006 and 2007} shows the ROUGE scores of related methods. The bolded results highlight the best results in the set of experiments. From the results, we have the following observstions:

1. Our proposed method (QE-WMVT+Mani) outperforms other related methods generally. Random and Lead sentences show the poorest ROUGE scores.

2. The effect of deep learning method is not very prominent in the query-oriented multi-document summarization, only slightly prominent in the score of ROUGE-2, and significantly lower in the score of ROUGE-1 than our proposed method. This shows that the method of deep learning needs to be improved.

3. Manifold ranking models based on query sentence expansion (QE-WMVT + Mani) have improved compared with the manifold ranking model MultiMR, JMFMR, JTMMR, JLTMMR, JLTMMR+SF and RDRP\_AP. This shows that the manifold ranking based on the expansion of the query sentence is generally better than the manifold ranking based on the original query sentence.

4. TextRank with TextRank expansion(QE-T+TextRank) can improve the effect of TextRank, but the overall performance is not as good as our method. This is because the TextRank method of the original query sentence is significantly worse than manifold ranking of the original query sentence (see result in section 5.1) and the query expansion method has limited ability to improve the model.

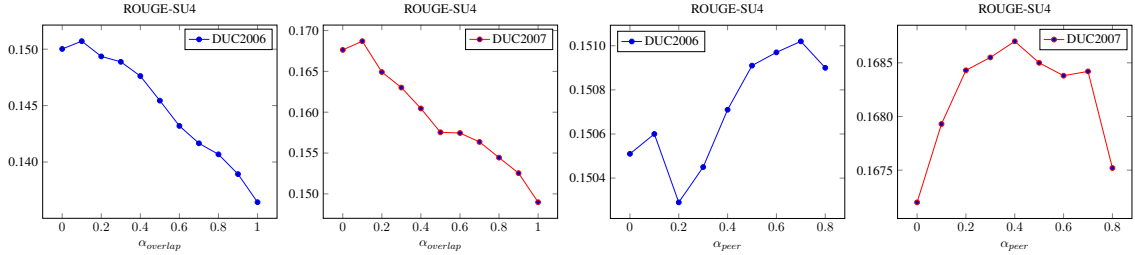
\begin{figure}[t]
	\begin{tikzpicture}[scale=0.45]
	\begin{axis}[
	legend columns=-1,
	legend entries={DUC2006},
	title={ROUGE-SU4},
	yticklabels={{},{},{0.140},{0.145},{0.150}},
	xlabel={$\alpha _{overlap}$},
	]
	\addplot+[sharp plot]               
	coordinates                         
	{                                  
		(0,0.150020)(0.1,0.15071)(0.2,0.14936)
		(0.3,0.14888)(0.4,0.14762)(0.5,0.14544)
		(0.6,0.1432)(0.7,0.14166)(0.8,0.14068)
		(0.9,0.13893)(1.0,0.13644)
	};
	\end{axis}
	\end{tikzpicture}
	\begin{tikzpicture}[scale=0.45]
	\begin{axis}[
	legend columns=-1,
	legend entries={DUC2007},
	title={ROUGE-SU4},
	yticklabels={{},{},{0.150},{0.155},{0.160},{0.165},{0.170}},
	xlabel={$\alpha _{overlap}$},
	]
	\addplot+[sharp plot,draw=red]               
	coordinates                         
	{                                  
		(0.0,0.16763) (0.1,0.16870) (0.2,0.16492) (0.3,0.16302) (0.4,0.16046) (0.5,0.15754) (0.6,0.15745) (0.7,0.15636) (0.8,0.15444) (0.9,0.15254) (1.0,0.14898)
	};
	\end{axis}
	\end{tikzpicture}
	\begin{tikzpicture}[scale=0.45]
	\begin{axis}[title={ROUGE-SU4},xlabel={$\alpha_{peer}$},
	legend entries={DUC2006},
	legend pos=north west,
	yticklabels={{0.1500},{0.1502},{0.1504},{0.1506},{0.1508},{0.1510}},
	]
	\addplot+[sharp plot] 
	coordinates               
	{                                  
		(0,0.15051)(0.1,0.1506)(0.2,0.15029)(0.3,0.15045)(0.4,0.15071)(0.5,0.15091)(0.6,0.15097)(0.7,0.15102)(0.8,0.1509)	
	};	
	\end{axis}
	\end{tikzpicture}
	\begin{tikzpicture}[scale=0.45]
	\begin{axis}[title={ROUGE-SU4},xlabel={$\alpha_{peer}$},legend entries={DUC2007},
	yticklabels={{},{},{0.1675},{0.1680},{0.1685}},
	]
	\addplot+[sharp plot,draw=red] 
	coordinates               
	{                                  
		(0,0.1672)(0.1,0.16793)(0.2,0.16843)(0.3,0.16855)(0.4,0.1687)(0.5,0.1685)(0.6,0.16838)(0.7,0.16842)(0.8,0.16752)	
	};
	
	\end{axis}
	\end{tikzpicture}
	\caption{ROUGE-SU4 scores vs. $\alpha _{overlap}$ ($\alpha _{overlap} + \alpha_A = 1$) and $\alpha _{peer}$.  }\label{fig:overlap_rate}
\end{figure}

\subsection{Influence of Sentence Similarity Parameter Tuning}
In our method, $\alpha _{overlap}$ is used to tune the trade-off between the TF-ISF and word overlap feature. $\alpha _{peer}$ is utilized to add the context feature. We carry out experiments with different $\alpha _{overlap}$ and $\alpha _{peer}$ to see their influence. 

Fig.\ref{fig:overlap_rate} presents the ROUGE-SU4 evaluation results of our method on DUC2006 and DUC2007 respectively for different values of $\alpha _{overlap}$ and $\alpha _{peer}$. In general, with an increase in $\alpha _{overlap}$ and $\alpha _{peer}$, the performance first increases, reaches its peak value and then is degraded. For DUC2006 and DUC2007, a value of $\alpha _{overlap}$ around 0.1 can attain the best ROUGE-SU4 result. This indicates that TF-ISF is a critical feature in sentence similarity, but word overlap feature can improve the effect of the model. For the DUC2006 dataset, a value of $\alpha _{peer}$ around 0.7 can attain the best ROUGE-SU4 result. For DUC2007,  $\alpha _{peer}=0.4$ produces peak values of ROUGE scores. Although the best value of  $\alpha _{peer}$ differ for DUC2006 and DUC2007, they can slightly improve the model results. This indicates that there are positional connections between sentences. 

\subsection{Query Relevance Performance}
We also perform the qualitative analysis to exam the query relevance performance of our model. We randomly choose some queries in the test datasets and calculate the relevance scores of sentences from our model. We then extract the top ranked sentences and check whether they are able to meet the query need. Examples for query and model summaries are shown in Table \ref{Query Relevance}. We also give the sentences from our model without query expansion for comparison.

With manual inspection, we find that most query-focused sentences in our model with query expansion can answer the query to a large extent. All these aspects are mentioned in reference summaries. The sentences from our model without query expansion, however, are usually short and simply repeat the key words in the query. The advantage of query expansion is apparent in query relevance ranking.

Different from single document summary, multi-document summary has more information, and the central idea of each document under each topic is different. Therefore, the amount of valid information, such as ROUGE, is more suitable for evaluating the effect of multi-document summarization. Although our model is based on extraction method, it is simple and efficient, and the amount of valid information can be comparable to the state-of-the-art models.

\begin{table}[t]
	\centering
	\begin{tabular}{|p{0.9\columnwidth}|}
		\hline
		\textsl{Query:} What has been the argument in favor of a line item veto? How has it been used? How have US courts, especially the Supreme Court, ruled on its constitutionality?\\
		\hline
		\textsl{Our model without query expansion Summary:} \\
		Supreme Court to Rule on Line Item Veto.\\
		Court rules Line Item Veto unconstitutional.\\
		Court strikes down president's Line Item Veto power.\\
		Supreme Court Rules Line Item Veto Law Unconstitutional.\\
		Clinton Vetoes Congress Rejection of His Line Item Veto.
		\\
		\hline
		\textsl{QE-WMVT+Mani Summary:} Supreme Court to Rule on Line Item Veto.\\
		In a 6-3 decision the high court ruled the Line Item Veto Act violated the constitution's separation of powers between Congress which approves legislation and the president who either signs it into law or vetoes it.\\
		Court strikes down president's Line Item Veto power.\\
		Clinton is the first president to have line item veto authority.\\
		In a 6-3 ruling the court said the 1996 Line Item Veto Act unconstitutionally allows the president by altering a bill after its passage to create a law that was not voted on by either house of Congress.\\
		\hline
	\end{tabular}
	\caption{Sentences recognized to focus on the query.}
	\label{Query Relevance}
\end{table}

\section{Conclusion}
\label{sec:length}

In this paper, we propose a query-oriented multi-document summarization model based on query expansion and manifold ranking. We use WordNet, mean, variance, TextRank to expand query. We also use the cosine values of TF-ISF, word overlap values and the relative position of sentences to calculate the relation between sentences. We applied our model to DUC2006 and DUC2007, and found that our model is better than one that combines a subset of query expansion methods and achieves competitive performance with the state-of-the-art results. 

\bibliography{mybibfile}

\end{document}